\documentclass{article}

\usepackage{arxiv}

\usepackage[utf8]{inputenc} 
\usepackage[T1]{fontenc}    
\usepackage{hyperref}       
\usepackage{url}            
\usepackage{booktabs}       
\usepackage{amsfonts}       
\usepackage{nicefrac}       
\usepackage{microtype}      
\usepackage{lipsum}
\usepackage{graphicx}

\title{SE-MelGAN - Speaker Agnostic Rapid Speech Enhancement}

\author{
  Luka Chkhetiani* \\
  Department of Machine Intelligence\\
  SYSTEM CORP.\\
  Tbilisi, Georgia 0160 \\
  \texttt{lchkhetiani@systemcorp.ai} \\
   \And
  Levan Bezhanidze* \\
  Department of Machine Intelligence\\
  SYSTEM CORP.\\
  Tbilisi, Georgia 0160 \\
  \texttt{lbezhanidze@systemcorp.ai} \\
}

\begin{document}
\maketitle

\begin{abstract}
{\it Recent advancement in Generative Adversarial Networks in speech synthesis domain \hyperlink{refer}{[3]}, \hyperlink{refer}{[2]} have shown, that it's possible to train GANs \hyperlink{refer}{[8]} in a reliable manner for high quality coherent waveform generation from mel-spectograms. We propose that it is possible to transfer the \textit{\textbf{MelGAN}}'s \hyperlink{refer}{[3]} robustness in learning speech features to speech enhancement and noise reduction domain without any model modification tasks. Our proposed method generalizes over multi-speaker speech dataset and is able to robustly handle unseen background noises during the inference. Also, we show that by increasing the batch size for this particular approach not only yields better speech results, but generalizes over multi-speaker dataset easily and leads to faster convergence. Additionally, it outperforms previous state of the art GAN approach for speech enhancement  \textit{\textbf{SEGAN}} \hyperlink{refer}{[5]} in two domains: 1. quality ; 2. speed. Proposed method runs at more than 100x faster than realtime on GPU and more than 2x faster than real time on CPU without any hardware optimization tasks, right at the speed of \textit{\textbf{MelGAN}} \hyperlink{refer}{[3]}. }
\end{abstract}

\keywords{MelGAN \and Multi-Band Melgan  \and GAN \and SEGAN  }
{
*The author occupies undergraduate studies for now
}

\section{Introduction}
Recent advancement of deep learning have pushed the frontiers of many fields. Performance and speed of speech synthesis methods have been increasing rapidly for the last few years. The revolutionary autoregressive \textit{\textbf{WaveNet}} \hyperlink{refer}{[4]} have been replaced with non-autoregressive, fully-parallel \textit{\textbf{Parallel WaveNet}} \hyperlink{refer}{[11]}, trained along with \textit{\textbf{WaveNet}} \hyperlink{refer}{[4]} with Probability Density Distillation objective for realtime speech synthesis. And lately proposed methods \textit{\textbf{MelGAN}} \hyperlink{refer}{[3]} and \textit{\textbf{Multi-Band Melgan}} \hyperlink{refer}{[2]} show that it is possible to handle mel-spectogram to raw waveform conversion at insanely high speeds, with comparable quality to previous state of the art methods.

After seeing the newest advancements in the speech synthesis domain, we came up with a logical question: Why cannot we use the robustness of speech vocoders in learning the speech features to handle noise reduction and speech enhancement problems? From one point of view, if a network is able to catch important features in mel-spectograms for high quality waveform generation - we can initially assume that they should be able to differentiate important speech features from noise in case they are properly trained.

\section{Problem Statement}

Some of the most distinguished researches  \hyperlink{refer}{[5]}, \hyperlink{refer}{[11]}, \hyperlink{refer}{[10]} have shown different methodologies for main speaker isolation from background or parallel noise. Face landmark-based speech enhancement technique proposed in \hyperlink{refer}{[10]} uses independent face landmark extractor  \hyperlink{refer}{[12]} model output as an input for LSTM-based networks to generate time-frequency masks, which later on are applied to mixed-speech spectrogram. However, for real-life situations there is rarely a case when we have main and background speech inputs on top of their frontal face video streams. 

\begin{flushleft}

Additionally, DLIB \hyperlink{refer}{[12]} framework is perfect for research purposes, but it suffers from landmark-lagging in real-time inference, along with the problem of ground-truth coordinates mismatch. And, there is no room for non-human background noise removal for this time.

However, VoiceFilter \hyperlink{refer}{[11]} uses targeted speaker's reference audio as a reference vector. Nevertheless the results are promising and in fact, the network significantly reduces Word Error Rate on noisy validation data in speech recognition setting, the inference time hardly reaches real-time factor of 0.5x on expensive NVIDIA Tesla V100 GPU.

On the other hand, SEGAN \hyperlink{refer}{[5]} promisingly laverages the background or parallel noise removal issue, but suffers from prominent speed problem, and is not able to nearly reach real-time factor even on expensive hardware.

Our aim is to find equilibrium, a network that is able to learn important speech features for isolation, and satisfy the need of excellent real-time inference factor on cheap hardware. The network should be lightweight and able to perform computations even locally with constrained memory budget in case the solution is served in a distributed manner, without additional hardware optimization tasks.

\end{flushleft}

\section{Methods}

\begin{flushleft}


One of the latest most exciting researches in mel-specrogram to raw audio conversion domain is non-autoregressive, fully feed-forward convolutional neural network \textbf{MelGAN} \hyperlink{refer}{[3]}. It is important to note, that the network consists of 4.26 million parameters, which is almost 21x less than the fastest available vocoder on the market \textbf{WaveGlow} \hyperlink{refer}{[13]}. Also a notable performance metric behind the \hyperlink{refer}{[3]} network is that while satisfying the inference speed and lightness requirement, the results are comparable to other heavy-weight networks with 3.09 MOS \textit{(Mean Opinion Score)}.

\end{flushleft}

\includegraphics[width=15cm]{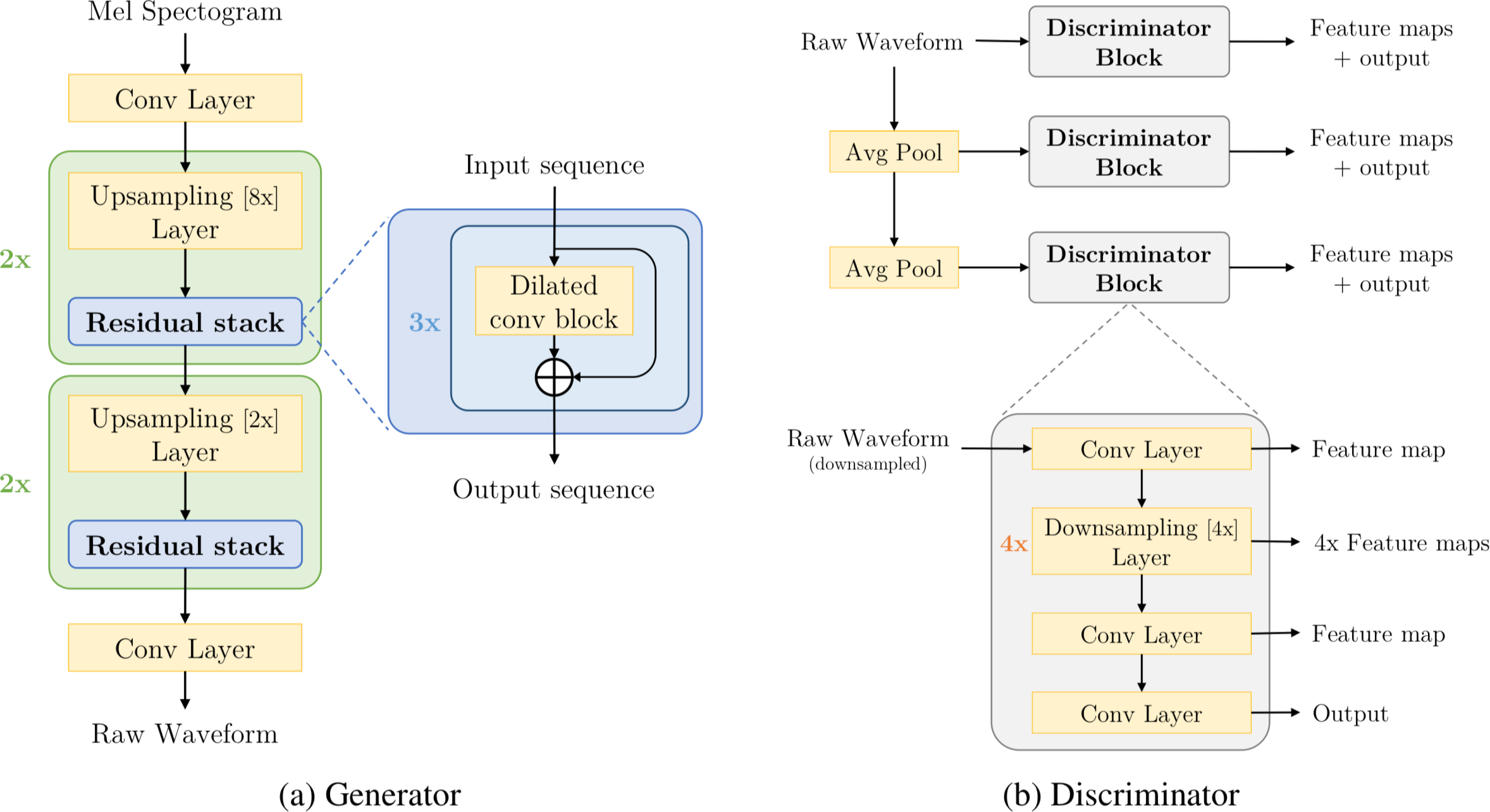}
\centering
\begin{center}
Figure 1. Proposed \textit{\textbf{MelGAN}} architecture in \hyperlink{refer}{[3]}. 
\end{center}

\begin{flushleft}

In our approach for speech enhancement we use the original version of proposed \textbf{MelGAN} architecture in \hyperlink{refer}{[3]}, but treat the training process as optimization of decompressing the reasonable speech features from mel-spectogram, rather than conversion of mel to raw audio in a straightforward manner.

\end{flushleft}

\begin{flushleft}

Our assumption is that as long as \textbf{MelGAN} \hyperlink{refer}{[3]} is almost perfectly able to detect and learn speech features in its receptive field, we should orient on training the model specifically for speech enhancement task, rather than start searching for perfect network modifications such as: increasing the receptive field, adding residual blocks for searching deeper features, etc. 
We use L2 loss as described in official paper, decrease the LeakyRelu activation slope from 0.2 to 0.01 and restrict the mel max-frequency to 8000hz. As long as most of the vocoders are able to learn speech features in restricted 0-8kHz frequency area, we assume that by increasing it model will end up tackling with more unnecessary noisy features that will lead to not-generalization and quality loss.
\end{flushleft}
\begin{flushleft}

We keep mel channels to 80, increase segment length to 16K. Filter, hop and window lengths are 1024, 256 and 1024 respectively, whereas sampling rate is set to 22050.

\end{flushleft}
\begin{flushleft}

Additionally, it's important to keep the input audio preprocessing in a discrete time, for which STFT \textit{(Short-Time Fourier Transform)} is used. By using convolutional STFT with 1D convolutions, the preprocessing time could be further reduced. 

\end{flushleft}

\section{Dataset}

\begin{flushleft}
For synthetic dataset generation we use LibriSpeech's \hyperlink{refer}{[9]} 360 hours version of multi-speaker speech data, along with \textbf{MS-SNSD} \hyperlink{refer}{[6]} and \textbf{ESC-50} \hyperlink{refer}{[7]} noises dataset. 

Objective for data generation is to increase the original data by 4x. 
Firstly, we randomly select clean speech audio sample from LibriSpeech \hyperlink{refer}{[9]} dataset. 
Secondly, we randomly select noise audio sample from combination of above-mentioned noisy datasets, and try to align them with clean speech. In case the aligned noise is shorter than speech, that happens most of the time, we cut most noisy parts iteratively and concatenate them so the final noise is exactly the same length as clean speech. The same is done in case the noise sample is longer than clean speech sample. 
Finally, we stack them together and perform short time fourier transformation with 1D Convolutions.  After tweaking the sample generated outputs, we have seen that most of the noisy speeches were non-audible. Thus we decrease the noisy sample volume by 30 percent, so that the final combination has audible speech in it.

Additionally, half of the clean dataset is used without augmentation. We assume that having a part of clean speech in training data will help model focus on important features of speech and easily converge.
\end{flushleft}

\begin{flushleft}
\textbf{ESC-50} \hyperlink{refer}{[7]} environmental noises dataset is as in the following table. Additionally, airport announcement, vacuum cleaner and neighbor speaking noise samples are used from \textbf{MS-SNSD} \hyperlink{refer}{[6]} dataset.
\end{flushleft}

\resizebox{\textwidth}{!}{%
\begin{tabular}{|l|l|c|c|c|}
\hline
\textbf{Animals} & \textbf{Natural soundscapes, water sounds} & \multicolumn{1}{l|}{\textbf{Human, non-speech sounds}} & \multicolumn{1}{l|}{\textbf{Interior/domestic sounds}} & \multicolumn{1}{l|}{\textbf{Exterior/urban noises}} \\ \hline
Dog                                & Rain                     & Crying baby                                & Door knock                                  & Helicopter                             \\ \hline

Rooster                                & Sea waves                     & Sneezing                                 & Mouse click                                  & Chainsaw                             \\ \hline
Pig                                & Crackling fire                     & Clapping                                 & Keyboard typing                                  & Siren                             \\ \hline
Cow                                & Crickets                     &  Breathing                                & Door, wood creaks                                  & Car horn                             \\ \hline
Frog                                & Chirping birds                      & Coughing                                 & Can opening                                   & Engine                             \\ \hline
Cat                                & Water drops                     & Footsteps                                 & Washing machine                                  & Train                             \\ \hline
Hen                                & Wind                     & Laughing                                 & Vacuum cleaner                                  & Church bells                             \\ \hline
Insects                                & Pouring water                     & Brushing teeth                                 & Clock alarm                                  & Airplane                             \\ \hline
Sheep                                & Toilet flush                     & Snoring                                 & Clock tick                                  & Fireworks                             \\ \hline
Crow                                & Thunderstorm                     & Drinking, sipping                                 & Glass breaking                                  & Hand saw                             \\ \hline
\end{tabular}%
}

\section{Training}
\begin{flushleft}

The training procedure is done on a server with two NVIDIA Tesla V100 GPUs with total of 32GB of VRAM. High-VRAM support gives us ability to handle bigger batch sizes, that as mentioned above, leads the model to converge faster and generalize well. We use batch size of 128 for training. As an optimizer, Adam with betas 0.5, 0.9 is used. 

Initially, we train the network for 1.5 Million steps with learning rate 1e-4, and decrease it to 1e-5 for further training for total of 3 Million steps.
\end{flushleft}

\begin{flushleft}
During experimentation, we found out that decreasing learning rate in the middle of training also helps the network to perform better and generalize on unseen noise samples and speakers well.

\end{flushleft}

\section{Results}

\begin{flushleft}

Output samples during manual validation process were chosen randomly.
In the noise-mixed samples we found several ones with multi or single speaker background voice over clean speech sample with noticeably high volume. In contrary with our initial idea, we saw that the network was able to not only differentiate speaker from random background noises but it started to choose main speaker over high-volume background speech and perfectly isolate it. In harsh terms, network generalized over reasonable speech features to isolate them, rather than on noises to take them away.

\end{flushleft}

\section{Conclusion}
\begin{flushleft}
We proposed a novel approach for training \textbf{MelGAN}\hyperlink{refer}{[3]}  for speech enhancement domain by using almost exactly the same network architecture as proposed in \hyperlink{refer}{[3]}. Results show, that nevertheless the network is designed specifically for raw waveform synthesis task, by carefully adopting it to speech enhancement issue we can yield comparable results to other state-of-the-art approaches while maintaining insanely fast inference speed.
Perhaps there's a much bigger room for adoption of SOTA networks to other domains than it was previously thought.

The implementation for non-commercial use will be available at: \href{https://github.com/systemcorp-ai/SE-MelGAN}{https://github.com/systemcorp-ai/SE-MelGAN} shortly.
\end{flushleft}
\begin{flushleft}
For audio samples, please visit: \href{https://systemcorp-ai.github.io}{https://systemcorp-ai.github.io}
\end{flushleft}

\section{Acknowledgements}
\begin{flushleft}
The authors would like to thank SYSTEM CORP. for funding and Seung-Won Park for amazing insights in their \textbf{MelGAN}\hyperlink{refer}{[3]} implementation repository, that gave us a very valuable help.
\end{flushleft}

\bibliographystyle{unsrt}  
\bibliography{references}
\hypertarget{refer}{}
\nocite{*}

\end{document}